\documentclass{iau_FM}
\usepackage{graphicx}

\newcommand{\kms}{\,{\rm km\,s^{-1}}}

\newcommand{\kpc}{\,{\rm kpc}}
\newcommand{\pc}{\,{\rm pc}}

\newcommand{\Myr}{\,{\rm Myr}}

\title[New Insights on Galactic Dynamos] 
{New Insights on Galactic Dynamos}

\author[Luke Chamandy]   
{Luke Chamandy$^{1}$\thanks{lchamandy@pas.rochester.edu}, 
Anvar Shukurov$^{2}$\thanks{anvar.shukurov@ncl.ac.uk} 
\& A.~Russ~Taylor$^{3,4}$\thanks{russ@ast.uct.ac.za}
}

\affiliation{
$^{1}$Department of Physics and Astronomy, University of Rochester, Rochester NY 14627, USA\\
$^{2}$School of Mathematics, Statistics \& Physics, Newcastle University, Newcastle~upon~Tyne NE1~7RU\\
$^{3}$Department of Physics and Astronomy, University of the Western Cape, Belleville 7535, Republic~of~South~Africa\\
$^{4}$Astronomy Department, University of Cape Town, Rondebosch 7701, Republic~of~South~Africa
}

\pubyear{2018}
\jname{Astronomy in Focus, Volume 8} 
\editors{???, ed.}
\begin{document}

\maketitle

\begin{abstract}
We argue that pitch angles of the azimuthally averaged large-scale or mean magnetic fields 
in nearby spiral galaxies inferred from observations can tentatively be explained with simple galactic dynamo models.
Agreement is not perfect, but is reasonable considering the uncertainty in dynamo parameters.
\end{abstract}

\firstsection 
\section{Introduction}
Large-scale (or mean) galactic magnetic fields are observed to be in near-equipartition
with turbulent kinetic energy in nearby galaxies, which suggests that the dynamo has likely reached saturation.
The most basic properties of the mean magnetic field are its strength $B$ and direction;
the latter is characterized by the pitch angle $p=\arctan(B_r/B_\phi)$, with $-90^\circ<p\leq90^\circ$,
where $B_r$ and $B_\phi$ are the radial and azimuthal components in cylindrical geometry $(r,\phi,z)$.
The quantity $p$ is arguably more useful for testing galactic dynamo theory than $B$, for three reasons.
Firstly, unlike $B$, $p$ is insensitive to the extent to which the dynamo instability is supercritical. 
Secondly, $p$ is insensitive to details of the dynamo non-linearity while $B$ is not.
And thirdly, $p$ is more directly and accurately inferred from observation than $B$.

The analytical expression for the magnetic pitch angle in the saturated (steady) state 
in the local axisymmetric dynamo model of \cite[Chamandy, Shukurov \& Subramanian (2014)]{Chamandy+14}
can be compared with pitch angles inferred from observations (\cite[Van Eck 2015 and references therein]{Vaneck+15}).
Because estimates of the effects of a mean outflow (wind or fountain flow) tend to be too small to significantly affect the dynamo
for the galaxies for which data exists to enable such estimates, the formula effectively reduces to
\begin{equation}
  \label{p_anal}
  p \simeq -\arctan\left[\frac{\pi^2\tau}{12q\Omega}\left(\frac{u}{h}\right)^2\right],
\end{equation}
where $\tau$ is the turbulent correlation time, $\Omega$ is the angular rotation speed, $q= -\mathrm{d}\ln\Omega/\mathrm{d}\ln r$
with $r$ the galactocentric distance ($q=1$ for a flat rotation curve), $u$ is the turbulent speed of the largest eddies
and $h$ is the scale height of diffuse gas. $p<0$ for a trailing spiral.
\cite[Van Eck et al. (2015)]{Vaneck+15} found that (i) theoretical values of $|p|$ are much too small compared with observations,
and (ii) observational and theoretical values are uncorrelated.
In \cite[Chamandy, Shukurov \& Taylor (2016)]{Chamandy+16} we refined their model in three ways: 
(i) we solved the $\alpha^2\Omega$ local axisymmetric dynamo equations numerically in 1.5D in $z$, 
evolving the mean magnetic field up to the steady state,
(ii) rather than taking $h$ as a parameter, we modeled $h(r)$ based on the flared HI Milky Way disc model 
of \cite[Kalberla \& Dedes (2008)]{Kalberla+Dedes08} scaled to $h=400\pc$ at $r=8\kpc$ (\cite{Ruzmaikin+88})
and scaled radially by the ratio of $r_{25}$ to that of the Milky Way, 
and (iii) rather than adopting $\tau=10\Myr$, we took $\tau$ to be a free parameter:
this leads to a `best fit' value of $\tau\approx14\Myr$.
Like \cite[Van Eck et al. (2015)]{Vaneck+15} we adopted $u=10\kms$, comparable to the sound speed of the warm gas,
and used the same compiled observational data for $\Omega(r)$ and $q(r)$.

\section{Accuracy and applicability of local axisymmetric solutions}

\begin{figure}
\begin{center}
 \includegraphics[height=1.3in,clip=true,trim=0 6 0 6]{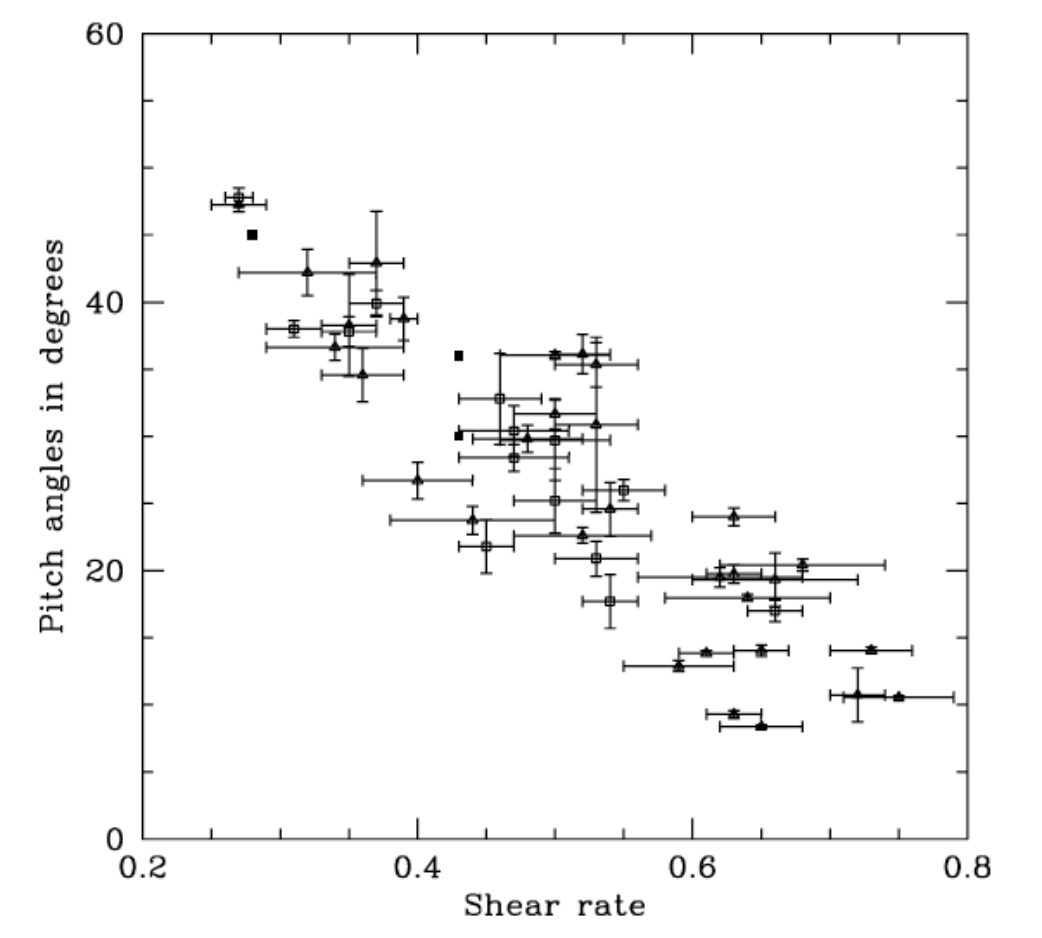} 
 \includegraphics[height=1.3in]{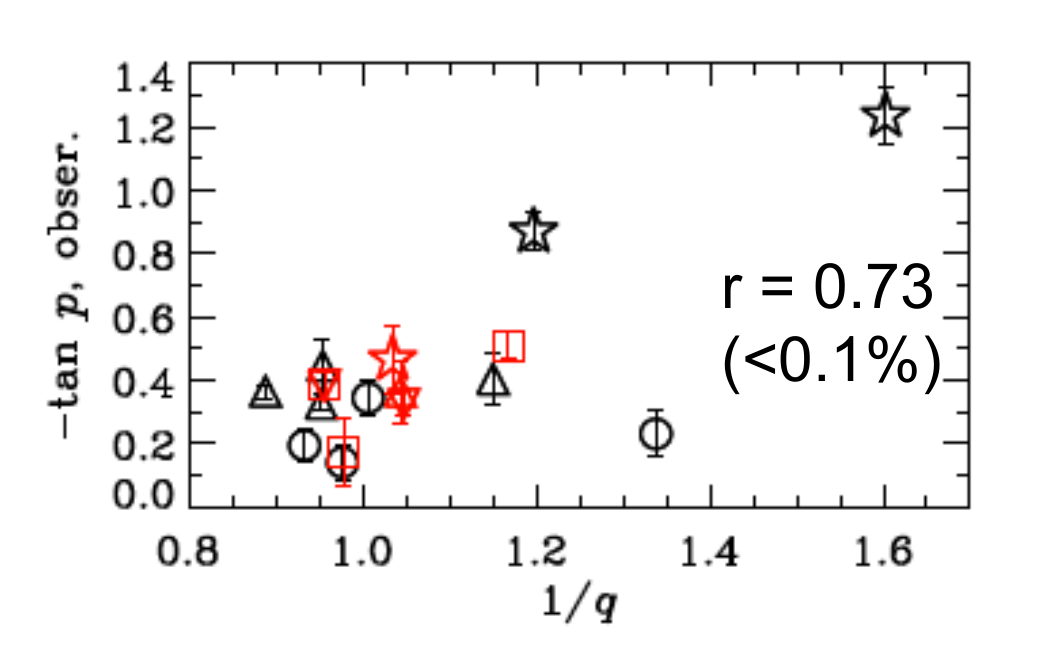} 
 \caption{Observational data showing correlations between $p_\mathrm{a}$ and $q$ (left, reproduced from \cite{Seigar+06}),
          and $p$ and $q$ (right, adapted from \cite{Chamandy+16}, with Pearson correlation coefficient and corresponding null probability).
          }
   \label{fig:shear}
\end{center}
\end{figure}

It has been shown that 1.5D axisymmetric saturated solutions and even simplified 0.5D solutions (see Eq.~\ref{p_anal}) 
approximate accurately 2.5D solutions in ($r$, $z$) outside of the central few hundred $\!\pc$ of the galaxy,
especially when galactic outflows are weak (\cite{Chamandy16}).
A possible objection to our approach is the neglect of non-axisymmetry.
While the $m=0$ azimuthal Fourier mode of the dynamo tends to dominate in axisymmetric discs, 
non-axisymmetric discs can excite higher order modes (e.g. \cite{Chamandy+15}),
but non-axisymmetric mean magnetic modes are generally found to be weak (\cite{Fletcher10}).
Weak non-axisymmetric modes are expected to exert only a very minor perturbation on the axisymmetric mode (\cite{Chamandy+13b}).
Nevertheless, it has been suggested that since the mean field tends to `follow' the spiral arms, there must be a causal effect.
Firstly, the premise is not supported by wavelet analysis which has shown that polarization angles in galaxies 
are consistently larger than spiral arm pitch angles $p_\mathrm{a}$ (\cite{Frick+16, Berkhuijsen+16, Mulcahy+17}).
Secondly, even if spiral arm and mean magnetic field pitch angles are correlated 
(whether this is true is not yet clear), such a correlation does not imply the existence of a causal relationship,
and, moreover, a correlation would indeed be expected from their common correlation with the shear rate.
This is demonstrated in Fig.~\ref{fig:shear}: the left panel shows a plot of $|p_\mathrm{a}|$ vs. $q/2$,
from \cite{Seigar+06}, while the right panel shows $\tan |p|$ vs. $1/q$ (see Eq.~\ref{p_anal})
from \cite[Chamandy, Shukurov \& Taylor (2016)]{Chamandy+16} 
(neglecting the galaxy M81 and SB-type galaxies, whose mean magnetic fields are observed to be highly non-axisymmetric;
see \cite{Krause+89} and \cite{Beck+05}).

\section{Results}

 \begin{figure}
 \begin{center}
  \includegraphics[width=5.2in]{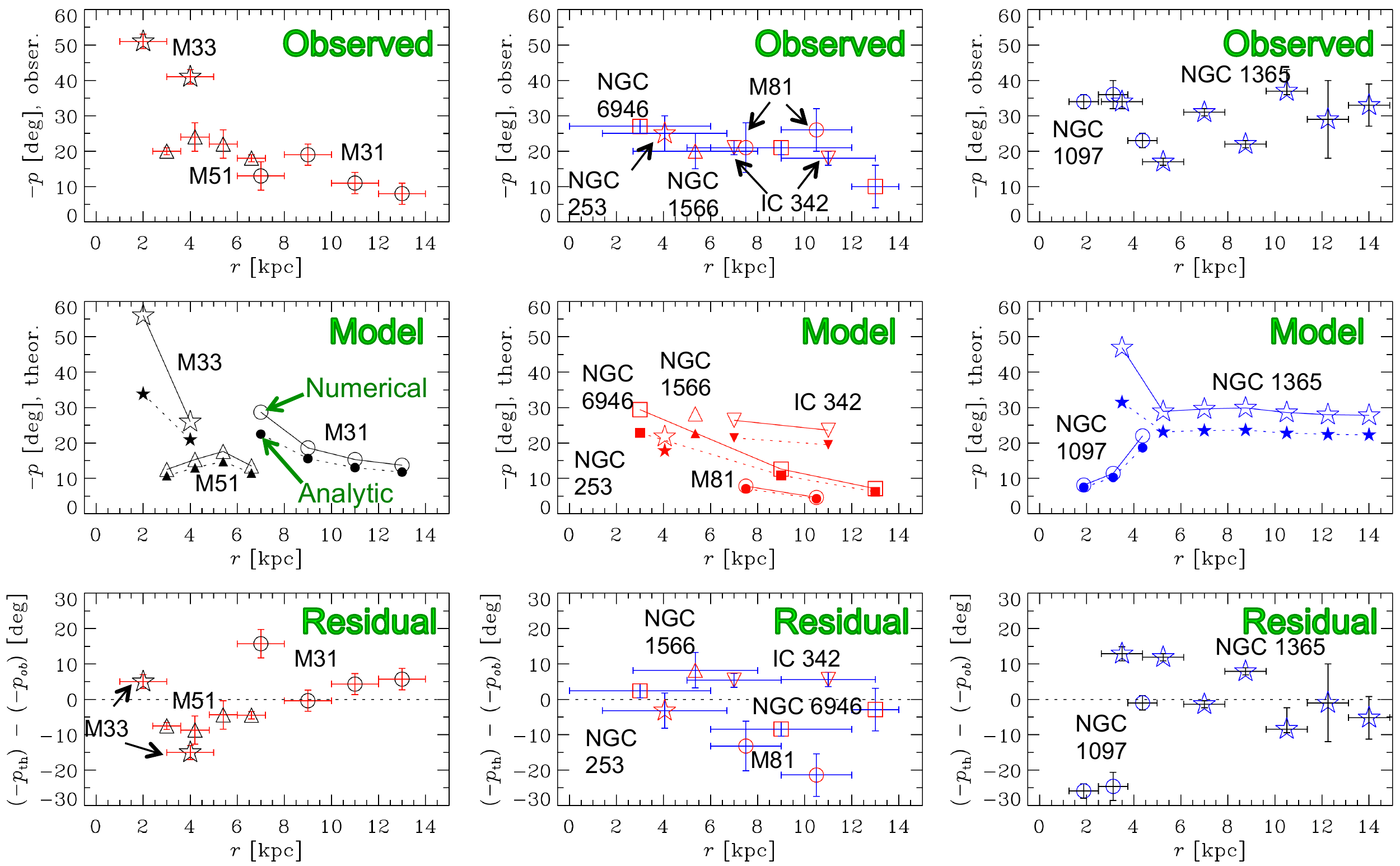} 
  \caption{Pitch angle data inferred from observations (top), obtained from our fiducial dynamo model (middle),
           and their difference (bottom) for three different types of galaxy, organized into columns 
           (adapted from \cite{Chamandy+16}, see text for details).}
    \label{fig:CST16_pitch}
 \end{center}
 \end{figure}
 
In Fig.~\ref{fig:CST16_pitch}, model results for $p$ 
(middle row, with open (closed) symbols representing numerical (analytical) solutions) 
are compared with values inferred from observations (top row), along with residuals (bottom row).
Galaxies are separated into columns according to their type and magnetic field observations:
the left-hand column contains SA galaxies for which Fourier analysis had been performed on the data
(these observational data are the most reliable of those available),
the middle column contains those SA or SAB galaxies for which such an analysis had not been performed,
and the right-hand column contains SB galaxies.
Agreement is not perfect, with $\chi_\nu^2\sim10$ for SA and SAB galaxies, 
but this is not surprising given that (i)~the model is idealized,
(ii)~inferred values rely on observational modeling and likely have systematic uncertainties, 
and (iii)~we chose to parameterize the model with only \textit{a single free parameter}, $\tau$,
even though other parameters are expected to vary between and within galaxies (\cite{Chamandy+Taylor15}).
For SB galaxies, which the model is \textit{not} meant to explain, 
the agreement is worse (especially at small radius where the bar is strong and the magnetic field highly non-axisymmetric)
than for unbarred or weakly barred galaxies.
Kendall's rank test can be performed on the pairs of pitch angles within each galaxy 
and shows that the model agrees with the SA and SAB galaxy data significantly better than a randomly ordered sample.

It is also encouraging that the `best fit' value of $\tau$ is of the expected order of magnitude.
We also tried a model with $h=\,$constant, i.e. an unflared disc, and found that, regardless of the value of $h$ used,
agreement between model and data was much worse than with our fiducial flared disc model.
We take this as evidence that galactic discs are flared, as would be expected from physical arguments
(\cite{Rodrigues+18}).
Galactic dynamo theory has also been used to point out that the alignment of magnetic spiral \textit{arms} 
(\cite{Beck+Hoernes96}) with their gaseous counterparts suggests that the latter are winding up and transient, 
as opposed to rigidly rotating and steady (\cite{Chamandy+15}).
Thus, magnetic fields can plausibly serve as probes of other phenomena, such as interstellar turbulence, 
disc flaring and spiral structure and evolution, even in cases where their dynamical influence may be weak or not fully apparent.

\begin{figure}
\begin{center}
 \includegraphics[height=1.4in]{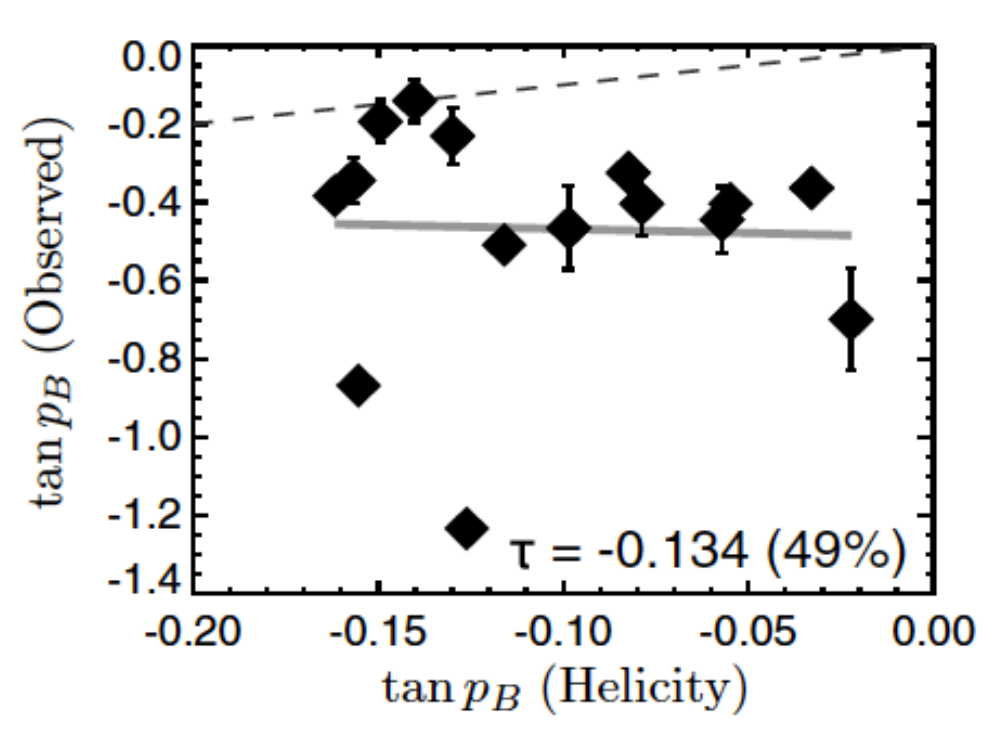} 
 \includegraphics[height=1.4in,clip=true,trim=0 0 0 20]{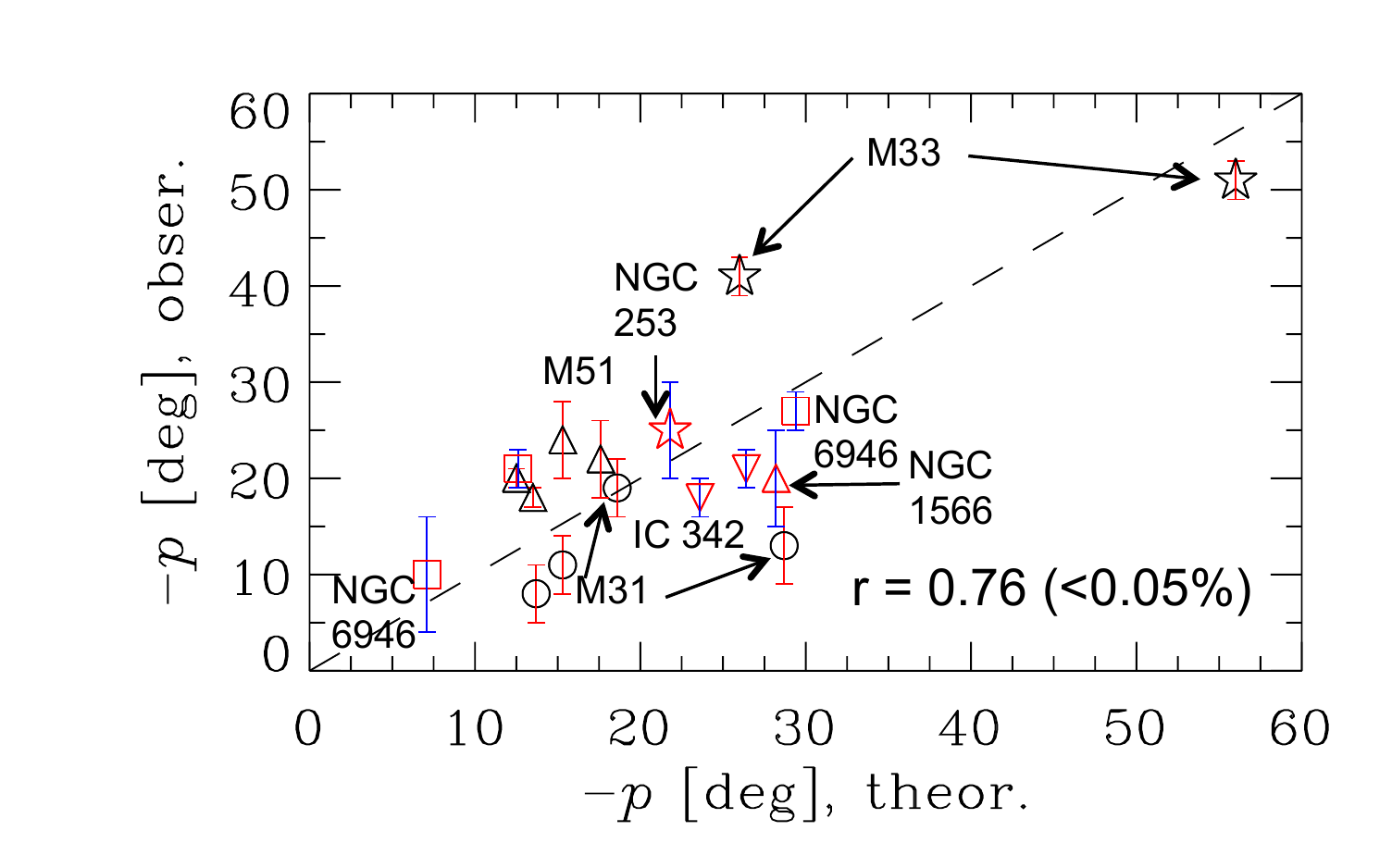} 
 \caption{Plot of pitch angles derived from observation vs. values obtained from theory for the models of \cite[Van Eck et al. (2015)]{Vaneck+15}
          (left, reproduced from that work) and \cite[Chamandy, Shukurov \& Taylor (2016)]{Chamandy+16} 
          (right, with Pearson correlation coefficient and corresponding null probability).}
   \label{fig:p_obs_theor}
\end{center}
\end{figure}

Finally, in Fig.~\ref{fig:p_obs_theor}, we compare our results (excluding M81 and SB galaxies, right panel) 
with those of \cite[Van Eck et al. (2015)]{Vaneck+15} (left panel), 
showing that the current state of affairs is not as dire if our more detailed model that includes disc flaring is invoked.

\section{Conclusions}
Our model can be made to be more realistic by including more physics: 
the gaseous galactic halo, 
independent constraints on parameters such as $u$, 
and revisiting the roles of outflows and non-axisymmetry.
However, given the paucity and heterogeneity of the data,
we feel that comparing with results of minimalistic dynamo models is a fruitful approach.
There is a need to analyze presently available data as uniformly as possible
with modern methods in order to expand and improve the data set.
At the same time, there is a need to come up with alternative dynamo models (e.g. \cite{Chamandy+Singh18}),
as well as to simulate magnetic properties for large populations of galaxies throughout cosmic time 
(\cite{Rodrigues+18}, see also L.~F.~S. Rodrigues, this proceedings).

\end{document}